\documentclass[preprints,article,accept,moreauthors,pdftex,10pt,a4paper]{Definitions/mdpi}
\usepackage[final]{changes}
\usepackage{tablefootnote}
\usepackage{upgreek}
\usepackage{textcomp}

\firstpage{1}
\pubvolume{xx}
\issuenum{1}
\articlenumber{5}
\pubyear{2019}
\copyrightyear{2019}
\history{Received: 30 January 2019; Accepted: 22 March 2019}
\def\mathLarge#1{\mbox{\Large $#1$}}

\Title{Faraday Rotation of Extended Emission as a Probe of the Large-Scale Galactic Magnetic Field}

\Author{Anna Ordog $^{1,*}$\orcidA{}, Rebecca A. Booth  $^{1}$, Cameron L. Van Eck $^{1,2}$\orcidC{}, Jo-Anne C. Brown $^{1,*}$\orcidD{} and Thomas L. Landecker $^{3}$}

\AuthorNames{Anna Ordog, Rebecca A. Booth, Cameron L. Van Eck, Jo-Anne C. Brown and Thomas L. Landecker}

\address{%
$^{1}$ \quad Department of Physics and Astronomy, University of Calgary, Calgary, AB T2N 1N4, Canada; rebecca.booth@ucalgary.ca\\
$^{2}$ \quad Dunlap Institute for Astronomy and Astrophysics, University of Toronto, 50 St. George Street, Toronto, ON~M5S~3H4, Canada; cameron.van.eck@dunlap.utoronto.ca\\
$^{3}$ \quad Dominion Radio Astrophysical Observatory, Herzberg Astronomy and Astrophysics Research Centre, National Research Council Canada, PO Box 248, Penticton, BC V2A 6J9, Canada; tom.landecker.drao@gmail.com\\}

\corres{Correspondence: aordog@ucalgary.ca (A.O.); jocat@ucalgary.ca (J.C.B.) }

\abstract{The Galactic magnetic field is an integral constituent of the interstellar medium (ISM), and knowledge of its structure is crucial to understanding Galactic dynamics. The Rotation Measures (RM) of extragalactic (EG) sources have been the basis of comprehensive Galactic magnetic field models. Polarised extended emission (XE) is also seen along lines of sight through the Galactic disk, and also displays the effects of Faraday rotation. Our aim is to investigate and understand the relationship between EG and XE RMs near the Galactic plane, and to determine how the XE RMs, a hitherto unused resource, can be used as a probe of the large-scale Galactic magnetic field. We used polarisation data from the Canadian Galactic Plane Survey (CGPS), observed near 1420 MHz with the Dominion Radio Astrophysical Observatory (DRAO) Synthesis Telescope. We calculated RMs from a linear fit to the polarisation angles as a function of wavelength squared in four frequency channels, for both the EG sources and the XE. Across the CGPS area, $55^{\circ} < {\ell} <193^{\circ}, -3^{\circ} < b < 5^{\circ}$, the RMs of the XE closely track the RMs of the EG sources, with XE RMs about half the value of
EG-source RMs. The exceptions are places where large local HII complexes heavily depolarise more distant emission. We conclude that there is valuable information in the XE RM dataset. The factor of 2 between the two types of RM values is close to that expected from a Burn
slab model of the ISM. This result indicates that, at least in the outer Galaxy, the EG and XE sources are likely probing similar depths, and that the Faraday rotating medium and the synchrotron emitting medium have similar variation with galactocentric distance.}

\keyword{galaxy: structure; ISM: magnetic fields; polarization; radio continuum: ISM; techniques: interferometric }

\begin{document}


\section{Introduction}
The idea that our Galaxy has a magnetic field was first proposed by Alfv\'en in 1937~\cite{Alfven}. Just over ten years later, Fermi corroborated this proposal, arguing that a Galactic magnetic field (\textbf{GMF}) would be able to explain the origin and confinement of cosmic rays. This field, he explained, would have coherence lengths on the order of thousands of \replaced{parsecs}{light years}, great stability, and would be able to prevent cosmic rays from escaping from the Galaxy~\cite{Fermi}. The first evidence for a large-scale GMF was from observations of starlight polarisation~\cite{Hiltner,Hall}.

The GMF is now considered to be a fundamental component of the interstellar medium (\textbf{ISM}). For example, magnetic fields may play a role in star formation~\cite{vanLoo}, and they contribute to the vertical support of gases in the Galaxy against gravitational collapse~\cite{Boulares}. Magnetic fields may also influence Galactic evolution~\cite{Kim}.

Most of what we know about the GMF has come from studying the Faraday rotation of polarised emission from compact, unresolved objects, such as extragalactic (\textbf{EG}) sources and pulsars~\cite{Simard, Rand, Brown2003b, Taylor2009}. Modeling the GMF has proven to be challenging, despite various approaches and use of a variety of datasets and assumptions~\cite{Jaffe, Van, Jansson, Gressel, Han}. Accounting for features of the GMF, such as the apparent large-scale reversal between the Local and Sagittarius arms~\cite{Simard, Van, Thomson} (for which no counterpart can be found in external galaxies), the variation in magnetic spiral pitch angle with galactocentric radius~\cite{Van}, and field components of varying spatial scales and degrees of coherence, adds complexity to the models being tested. The difficulty in constraining these models has been compounded by the sparse nature of the EG source data available.

Faraday rotation of polarised extended emission (\textbf{XE}) is another potential source of information on the GMF structure. While weaker in polarised intensity than the EG sources, XE occurs in nearly all directions, and its Faraday rotation may be useful in conjunction with EG sources. Highly polarised portions of the sky, for example, the Fan region, have been studied in terms of Faraday rotation of the diffuse emission~\cite{Iacobelli}, but very little investigation has been done into utilizing this technique for studying the large-scale GMF. In this paper we explore that possibility.

For compact sources, it is reasonable to assume that the polarised emission originates from a single point in space (i.e., that there is no significant internal Faraday rotation), and that the rotation occurs along a clearly identified path through the Galaxy. In this case, we calculate the Rotation Measure (\textbf{RM}) as the slope of a least-squares linear fit to the polarisation angle ($\tau$) as a function of the square of the wavelength ($\lambda^2$), so that
\begin{equation}
\tau = \tau_{\circ} + RM \lambda^2.
\label{eq:linear}
\end{equation}

\noindent This is subtly different in the case of XE, for which synchrotron emission and Faraday rotation occur within the same volume, both linked to the same magnetic field. In this case, we use the term Faraday depth to quantify the effects of Faraday rotation from a specific volume at a particular distance, $d$, from the observer located at the origin. Faraday depth is defined by the line integral:
\begin{equation}
\phi = 0.812\int_{d}^{0} n_e\, B_{\parallel}\,d\ell,
\label{eq:RM}
\end{equation} where $n_e$ is the electron density, $B_{\parallel}$ is the parallel component of the GMF, and $d\ell$ is the infinitesimal path length along the line-of-sight (\textbf{LOS})\added{, from the source to the observer}. For a Faraday-thin source without internal Faraday Rotation, as is typically assumed for a compact EG source, RM and Faraday Depth are identical. In contrast, the Faraday Depth profile of XE regions can be quite complicated~\cite{Burn, Sokoloff}, and it is generally argued that wider coverage and higher resolution in wavelength are required for studying Faraday Rotation of XE than for compact sources~\cite{Brentjens}.

In general, the RM of a polarised source provides information about the strength and the direction of the average magnetic field component parallel to the LOS, scaled by the electron density along the path. A positive [negative] RM indicates that the average LOS component of the GMF points toward [away from] the observer.

Recent discussions about the GMF, for example, by Ordog et al.~\cite{Ordog}, have expanded to use the RMs calculated from four frequency channels as a proxy for Faraday Depth of XE. From here forward we speak of the ``RM of XE'' with this specific meaning. The XE data used in~\cite{Ordog} were from the Dominion Radio Astrophysical Observatory (\textbf{DRAO}) Synthesis Telescope \textit{without} the addition of single-antenna data, that is, from interferometers of baselines 13 m to 600 m. It can be argued that XE data observed using radio interferometry are an incomplete source of information, particularly on polarisation angle and related quantities, due to a lack of sensitivity to  large-scale structures. However, Ordog et al.~\cite{Ordog} demonstrated a remarkable similarity between the RMs seen in the EG and the XE over a large area using this type of dataset. The reversal in magnetic field direction between the Local and Sagittarius arms, between longitudes 55$^{\circ}$ and 70$^{\circ}$, can be seen in the RM patterns of both the EG sources and the XE, occurring in the same place.

In exploring the relationship between the RMs of EG sources seen through the Galactic disk and the RMs or Faraday depths of XE, it would be ideal to have high-resolution data that include information on a wider range of spatial scales, achieved by combining aperture-synthesis and single-antenna data (e.g.,~\cite{Landecker}). However, we do not yet have single-antenna data with multiple frequency channels that can be combined with the synthesis telescope data. Thus we investigated the utility of the XE data from the DRAO Synthesis Telescope alone. In particular, we report on our efforts to learn as much as we can about the GMF by comparing the RMs of EG sources and XE from the Canadian Galactic Plane Survey (\textbf{CGPS}), with the aim of developing methods for combining information from the two types of datasets.

Our paper outline is as follows. In Section \ref{s2} we describe the datasets used. In Section \ref{s3} we compare the EG and XE sources, commenting on the observed trends. In Section \ref{s4} we discuss possible simple models of the ISM that could account for the correlations we find between the two types of sources and comment on anomalous regions. In Section \ref{s5} we summarize our results.

\section{The Data}\label{s2}

The CGPS is a radio survey centred at a frequency of 1420 MHz between Galactic latitudes $-3^{\circ} \le b \le  5^{\circ}$ with a Galactic longitude range of $52^{\circ} \le \ell \le 193^{\circ}$. \added{The survey specifications are listed in Table \ref{specs}.} From the polarisation data collected as part of the CGPS using the Synthesis Telescope at the DRAO, we have calculated RMs for XE across the entire dataset.

We derived RM values from the CGPS Stokes $Q$ and $U$ images in four bands, each 7.5 MHz wide, centred at the frequencies \replaced{indicated in Table \ref{specs}}{ 1407.2 MHz, 1414.1 MHz, 1427.7 MHz, and 1434.6 MHz} \cite{Landecker}. From the Stokes $Q$ and $U$ images we calculated the polarisation angle, $\tau$, for each frequency using $\tau = \frac{1}{2}\arctan U/Q$. The RM for each pixel in the image was determined using Equation~(\ref{eq:linear}). By using this algorithm, which is identical to that used in determining the EG source RMs (Van Eck et al. 2019, in prep., based on the method described in~\cite{Brown2003b}), we are treating the two datasets in the same manner. This process yielded $4.9\times 10^7$  RM values spanning the entire CGPS region, stored in an 1825 $\times$ 29,025 pixel array on a 20 $\times$ 20 arcsecond grid.

\begin{table}[H]
\caption[]{\added{The CGPS polarisation survey.}}
\label{specs}
\centering
\begin{tabular}{ll}
\toprule
Overall coverage             & ${53^{\circ}} < {\ell} < {192^{\circ}},
                               {-3^{\circ}} < {b} < {5^{\circ}}$ \\
                             & ${101^{\circ}} < {\ell} < {116^{\circ}},
                               {5.0^{\circ}} < {b} < {17^{\circ}}$ \\\midrule
Continuum bandwidth          & 30\,MHz in four bands of \\
                             & 7.5\,MHz each \\\midrule
Polarisation products        & Stokes $I$, $Q$, and $U$ \\\midrule
Centre frequencies           & 1407.2, 1414.1, 1427.7,\\
                             & and 1434.6\,MHz \\\midrule
Angular resolution           & $58'' \times 58''$\,cosec\thinspace$\delta$ \\\midrule
Sensitivity, $I$             & 200 to 400 $\upmu$Jy/beam rms \\\midrule
Sensitivity, $Q$ and $U$     & 180 to 260 $\upmu$Jy/beam rms \\\midrule
Typical noise in             & \\
mosaicked images             &  ${76}\thinspace{\rm{sin{\thinspace}}}\delta$~mK \\\midrule
* Sources of                   & Effelsberg 100-m Telescope \\
single-antenna data        & and DRAO 26-m Telescope \\
\bottomrule
\end{tabular}
\begin{tabular}{@{}c@{}}
\multicolumn{1}{p{\textwidth -.88in}}{\footnotesize * These data are part of the full-band CGPS survey,\ but are not available for the individual 7.5 MHz bands. Therefore we do not use these single-antenna data in our analysis.}
\end{tabular}
\label{tab:surprop}
\end{table}
\unskip

\subsection{Data Processing}

\begin{figure}[H]
\centering
	\includegraphics[width=0.75\textwidth]{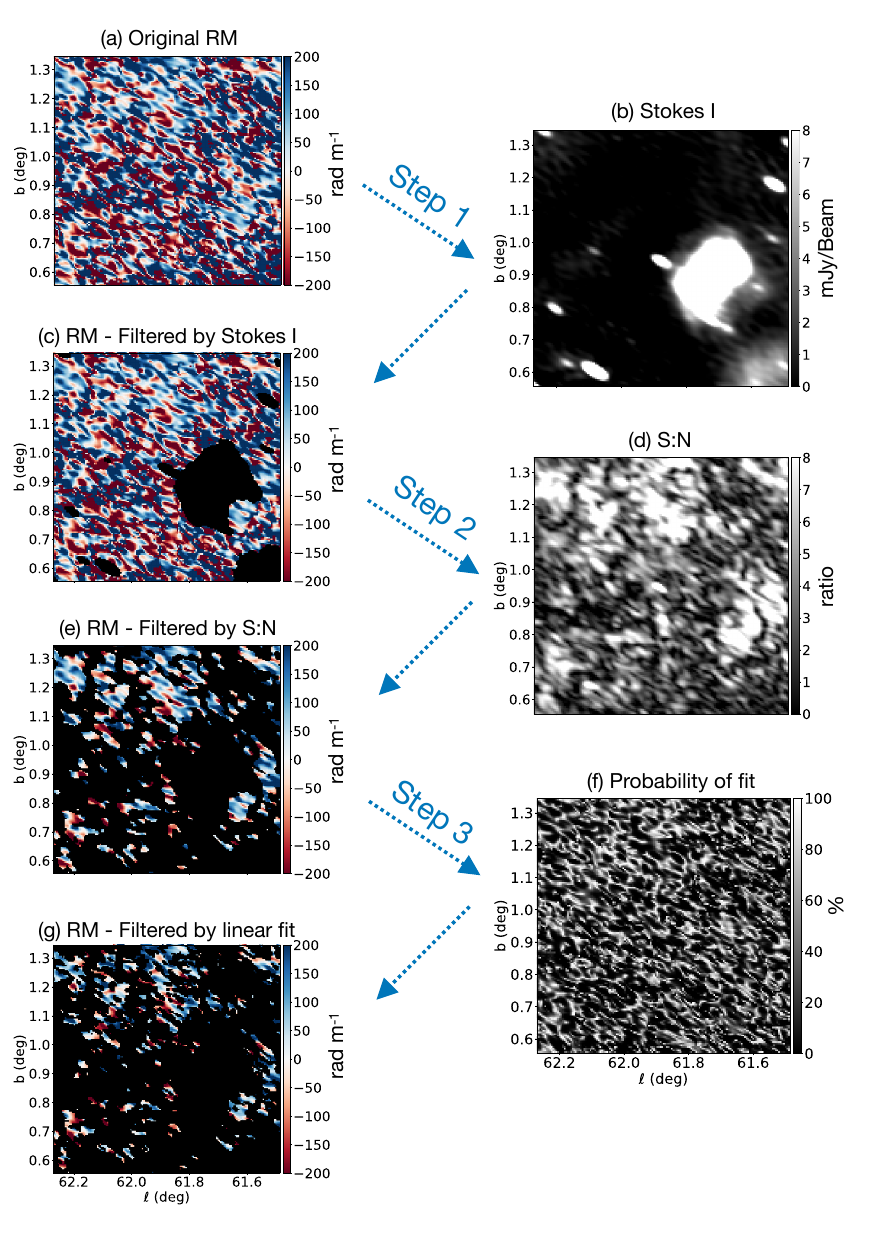}
   \caption{The process by which we filtered the XE RM data, illustrated for a selected region of the CGPS RM image. (\textbf{a}) The raw RM values from the linear fit to polarisation angle versus $\lambda^2$. (\textbf{b}) The corresponding Stokes $I$ image used in the first processing step. (\textbf{c}) The RM map with pixels with Stokes $I$ greater than 3 mJy/beam removed. (\textbf{d}) The signal-to-noise map, used in the second processing step. (\textbf{e}) The RM map with pixels in polarised intensity below the signal to noise threshold of 5 removed. (\textbf{f}) The probability of linear fit used in the third processing step. (\textbf{g}) The RM map with pixels with a probability of linear fit less than 10\% removed. }
\label{fig:data_processing}
    \end{figure}

In order to extract the XE information from the entire CGPS RM array and to ensure reliable RM data, we examined the corresponding Stokes $I$ image, the signal to noise (\textbf{S:N}) ratio of the \replaced{linearly polarised intensity, $P=\sqrt{Q^2+U^2}$}{polarisation data}, and the linear fits of angle as a function of $\lambda^2$ for each pixel. The process by which we filtered the XE RM dataset involved three steps as described below and illustrated in Figure~\ref{fig:data_processing}. \footnote{The final filtered XE RM map will be available at the CDS via http://vizier.u-strasbg.fr/viz-bin/VizieR.}

\unskip

\subsubsection*{\textit{Step 1: Remove pixels with high Stokes I}}

In addition to the extended synchrotron emission, in the \replaced{polarisation}{ polarised} images we also see compact sources and discrete objects such as supernova remnants and HII regions. While the HII regions do not emit polarised radiation, they can affect the polarised background. Both the compact objects and discrete objects tend to be brighter in Stokes $I$ than the surrounding XE, making it relatively straightforward to remove them from the XE RM images.  Removing these objects is necessary, since we are interested in comparing the XE RMs to the EG RMs. Figure~\ref{fig:data_processing}a--c demonstrate an example of this stage of the filtering for a small segment of the RM image.

In order to identify the RMs of XE only, we discarded all pixels from the RM array with corresponding Stokes $I$ higher than 3 mJy/beam. We experimented with a variety of thresholds and found that choosing a 3 mJy/beam threshold successfully removed all EG sources as well as pixels from known bright Galactic features from the RM data. After this 3 mJy/beam Stokes $I$ threshold was applied, $4.6\times 10^7$ pixels (94\% of the original number) remained in the RM array.

\subsubsection*{\textit{Step 2: Removal of pixels with low signal-to-noise ratio}}

Adopting the practice used in a previous work on the CGPS RM data~\cite{Brown2003b}, pixels with low S:N, for which the signal detected in polarised intensity was less than five times the noise, were rejected from the XE RM analysis, as shown in Figure~\ref{fig:data_processing}c--e. \added{We determined a S:N map based on the sensitivity of the Synthesis Telescope (see Table~\ref{tab:surprop}) and the weight maps of the images. The method is presented in detail in~\cite{Brown2003b} (Appendices 1 and 2) and~\cite{Brown2002}. We then used this S:N map as a filter for the XE RMs.} Following this step, $4.0\times 10^6$ RM pixels (8.2\% of the original number) remained in the array.

\subsubsection*{\textit{Step 3: Removal of pixels with fit probability < 10\%}}

In order to ensure reliable RM values, we determined the quality of the linear fit for each RM using a standard Pearson's chi-squared test that compares the calculated chi-squared values to a chi-squared distribution. We only accepted pixels with reliable linear fit parameters, rejecting those with fit probability less than 10\%, as shown in Figure~\ref{fig:data_processing}e--g. Following this filtering step, $2.3\times 10^6$ RM pixels (4.7\% of the original number) remained. Although a significant number of the pixels (55\% of the original) had a low probability of fit, more than 75\% of these were already discarded due to low S:N.

The pixels with poor linear fit likely correspond to lines of sight having increased Faraday complexity. With only four frequency channels, we are not able to accurately extract useful information for those lines of sight (e.g., using RM synthesis~\cite{Brentjens}). Therefore, we do not include them in this~analysis.

The catalog of EG RMs for the CGPS has a total of \replaced{2112}{ 2111} sources \added{in the Galactic disk}, resulting in close to 2 sources per square degree (Van Eck, et al., 2019, in prep.). However, to ensure that the EG sources and the XE used in our analysis were probing similar lines of sight, we did not include EG sources located along lines of sight passing through Stokes $I$ regions brighter than 3 mJy/beam, since those regions were already eliminated from the XE RM map. We identified such EG sources by their locations within the 3 mJy/beam threshold mask. Using this criterion, \replaced{2084}{ 2083} EG sources remained.

\section{Analysis}\label{s3}
\begin{figure}[H]
	\centering
    \includegraphics[width=1.0\textwidth]{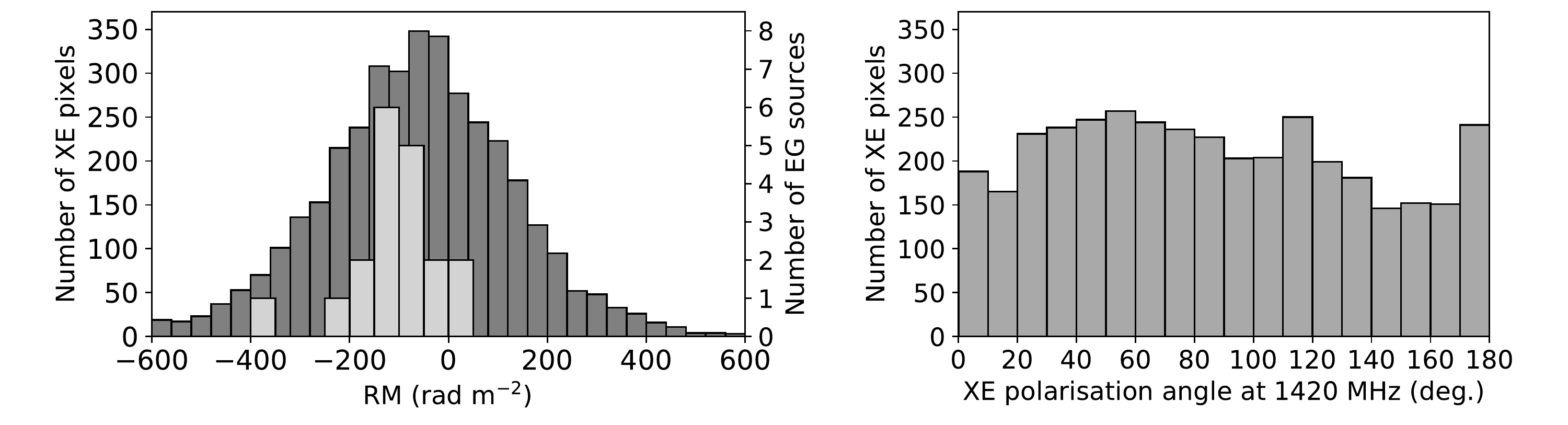}
   \caption{\added{\textbf{Left:} The distribution of XE RM pixels after the filtering process (dark gray) and EG RM sources (light gray) within one 1.5$^{\circ}$ wide longitude bin, centred at $\ell=137.5^{\circ}$. \textbf{Right:} The distribution of XE polarisation angles measured at 1420 MHz within the same longitude bin. While the XE polarisation angles seem to be randomly distributed within the bin, the XE RM values have a clear peak.}}
\label{fig:histograms}
    \end{figure}

We compared RM trends as a function of longitude in the XE and EG RM datasets by calculating the average for both within independent bins each spanning 1.5 degrees of longitude and covering the CGPS latitude range. \added{Figure~\ref{fig:histograms} shows an example of the distribution of XE and EG RMs within one of the bins. There is a high degree of scatter in the XE RM values, as can also be seen in the small-scale variability in the map shown in Figure~\ref{fig:data_processing}g. Nevertheless, the distributions within the bins do have clearly defined peaks. The use of sophisticated methods for producing a smoothed EG RM map, such as~\cite{Oppermann}, is beyond the scope of this paper. For the purpose of observing large-scale trends as a function of longitude, we have opted to simply average the EG RMs within each bin.}

\begin{figure}[H]
	\centering
    \includegraphics[width=1.0\textwidth]{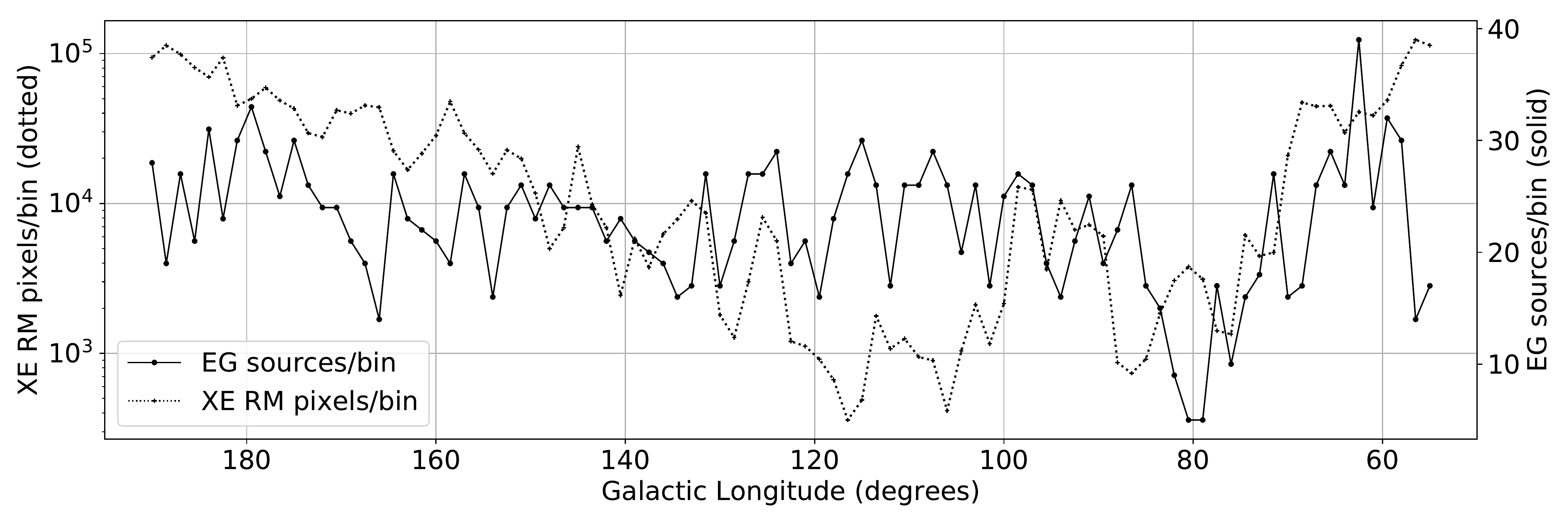}
   \caption{The number of XE RM pixels (dotted line) and EG sources (solid line) contained in independent bins each spanning 1.5 degrees of longitude and covering the CGPS latitudes, $b=-3^{\circ}$~to~$+5^{\circ}$.}
\label{fig:data_remaining}
    \end{figure}

As described above, the filtered XE RM array contains $2.3\times10^6$ pixels, 4.7\% of the original number. Even with this small fraction of the XE data remaining, the number of lines of sight probed by the XE still greatly exceeds the number probed by EG sources \footnote{The pixels are not independent, because the synthesized beam of the telescope consists of between 9 and 24 pixels in the image, depending on the declination. However, even with this taken into account, we still gain lines of sight in the XE that are missing in the EG source list.}. Figure~\ref{fig:data_remaining} shows the number of XE RM pixels and the number of EG sources in each 1.5 degree longitude bin. The 1.5 degree bin width was selected to ensure that there was a statistically significant number of EG sources contained in each bin. On average there are 22 EG sources in each bin with a low of \replaced{5}{ 4} sources/bin at \added{$\ell=79.0^{\circ}$ and} $\ell=80.5^{\circ}$. This is in the Cygnus X region, a local, extended region of thermal emission which depolarises emission produced behind it. We discuss this region further in Section \ref{s4}.

\unskip

The average number of XE RM pixels in each 1.5 degree bin is $2.3\times 10^4$. However, there is a decreasing trend towards the middle of the longitude range, with only \replaced{359}{ 353} pixels in the bin centred at $\ell=116.5^{\circ}$. Three effects contribute to this reduced number of pixels in central longitudes of the CGPS~region:

\begin{enumerate}
\item The intensity of synchrotron emission \replaced{depends on}{ is proportional to} the component of the magnetic field perpendicular to the LOS \added{, as well as the relativistic electron density in the emitting region}. \replaced{The intensity}{ This component} peaks \replaced{close to the}{ in the} anticentre\added{, in the Fan region (near $\ell=150^{\circ}$),} and falls smoothly toward lower \replaced{longitudes}{ declination}.
\item The synthesized beam of the telescope varies with declination, with a smaller beam having less signal for the same polarised brightness temperature. The beam is smallest at longitude about 120$^{\circ}$, \added{corresponding to the highest declination,} where there are fewer pixels in the beam to survive the various filters that we have applied.
\item The shortest baseline used by the Synthesis Telescope is 12.9 m. At low declinations foreshortening of the baseline extends the sensitivity of the telescope to larger angular structures. Conversely, at high declinations the sensitivity to extended structure is somewhat reduced. The highest declinations correspond to roughly $\ell = 120^{\circ}$. \added{Although it would be possible to make the sensitivity to extended structures uniform across the dataset by tapering the visibilities of long baselines, doing so would not yield significant gains, since it would correspond to an overall loss of high quality data.}
\end{enumerate}

\begin{figure}[H]
	\centering
    \includegraphics[width=1.0\textwidth]{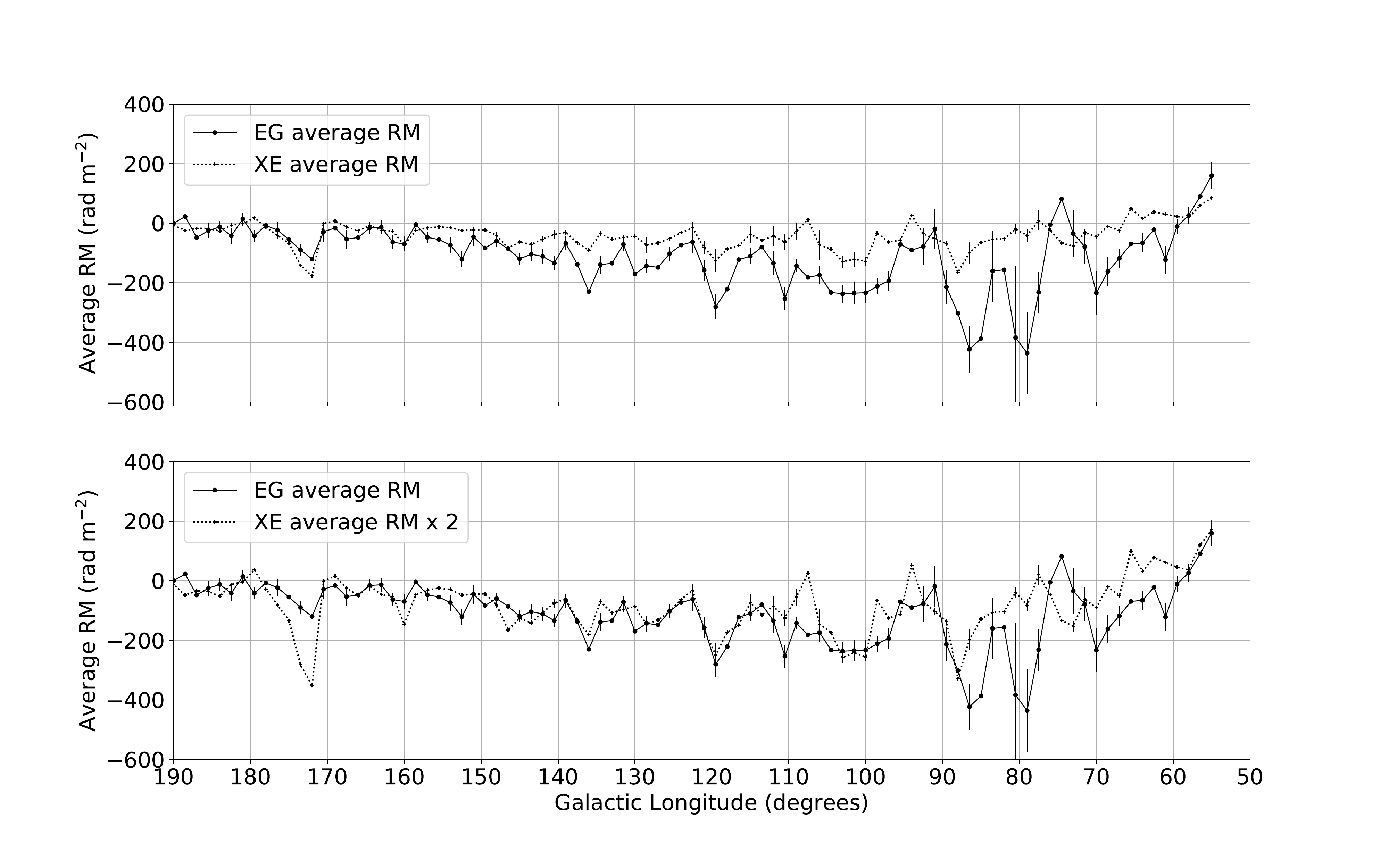}
   \caption{\textls[-10]{\textbf{Top:} The average EG RMs (solid line) and the average XE RMs (dotted line) within independent bins each spanning 1.5 degrees of longitude and covering the CGPS latitudes, $b = -3^{\circ}$ to $+5^{\circ}$. \textbf{Bottom:} The average EG RMs (solid line) and the average  XE RMs multiplied by two (dotted line).}}
\label{fig:averaging}
    \end{figure}

The top panel of Figure~\ref{fig:averaging} is a plot of the averaged RM in 1.5 degree slices in Galactic longitude for XE and EG, as function of Galactic longitude. The error bars for the EG RM bins are the standard deviation of the mean within the bins. The error bars for the XE RM bins were determined by dividing the standard deviation of the XE RMs within each bin by the square root of the average number of synthesized beams within the bin, to account for pixels not being independent. While the average XE RM is smaller in magnitude than the average EG RM, the most striking result is that the XE and EG RMs seem to be following the same trend with longitude. \added{The CGPS EG dataset has been shown to be consistent with other published datasets such as~\cite{Taylor2009,Van}, as demonstrated in Van Eck et al. 2019, in prep.}

\replaced{In order to evaluate the correlation between the binned XE and EG lists, we determined the Pearson correlation coefficient and found a result of}{ The Pearson correlation coefficient between the binned XE and EG lists is} \replaced{0.53}{ 0.52} (\emph{p} = 1$\times10^{-5}$), indicative of a proportionality between the two types of RM sources. We experimented with different bin widths, and with different values of the parameters in the XE data filtering process, and found that the results shown in Figure~\ref{fig:averaging} were robust with respect to our choice of these parameters.

Burn (1966) calculated the theoretical RM that would be observed for a LOS through a slab with uniform electron density, magnetic field, and synchrotron emission profiles~\cite{Burn}. In this simplified model, the Faraday Depths average along the LOS resulting in an observed RM for XE that is half that of the emission from an EG source along the same LOS. Adopting the Burn slab model as a starting point, we multiplied the XE RM average by two, as shown in the bottom panel of Figure~\ref{fig:averaging}, and were struck by how well the two datasets then line up. This result is surprising given that the magnetic field, electron density and synchrotron emissivity are not constant throughout the Galactic disk. Nonetheless, the agreement in overall trend and detail between the average XE RM and EG RM curves is undeniable, especially in the portion of the outer Galaxy spanned by the CGPS ($90^{\circ} \le \ell \le193^{\circ}$).

\begin{figure}[H]
	\centering
    \includegraphics[width=0.5\textwidth]{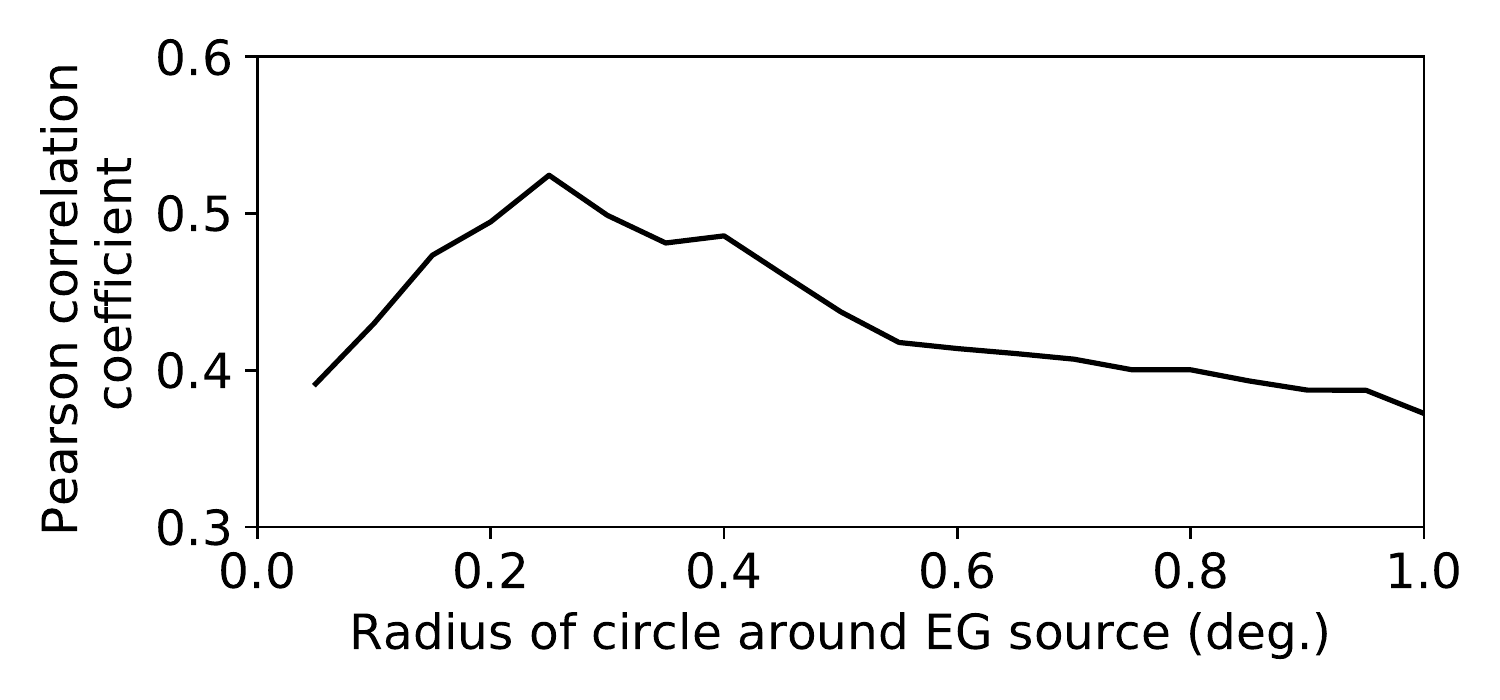}
   \caption{\added{The Pearson correlation coefficient between the XE RM values averaged within circles of varying radii centred at each EG source, and the corresponding EG RM value. The correlation is strongest for XE areas with a radius of 0.25$^{\circ}$.}}
\label{fig:correlation}
    \end{figure}

\begin{figure}[H]
	\centering
    \includegraphics[width=0.9\textwidth]{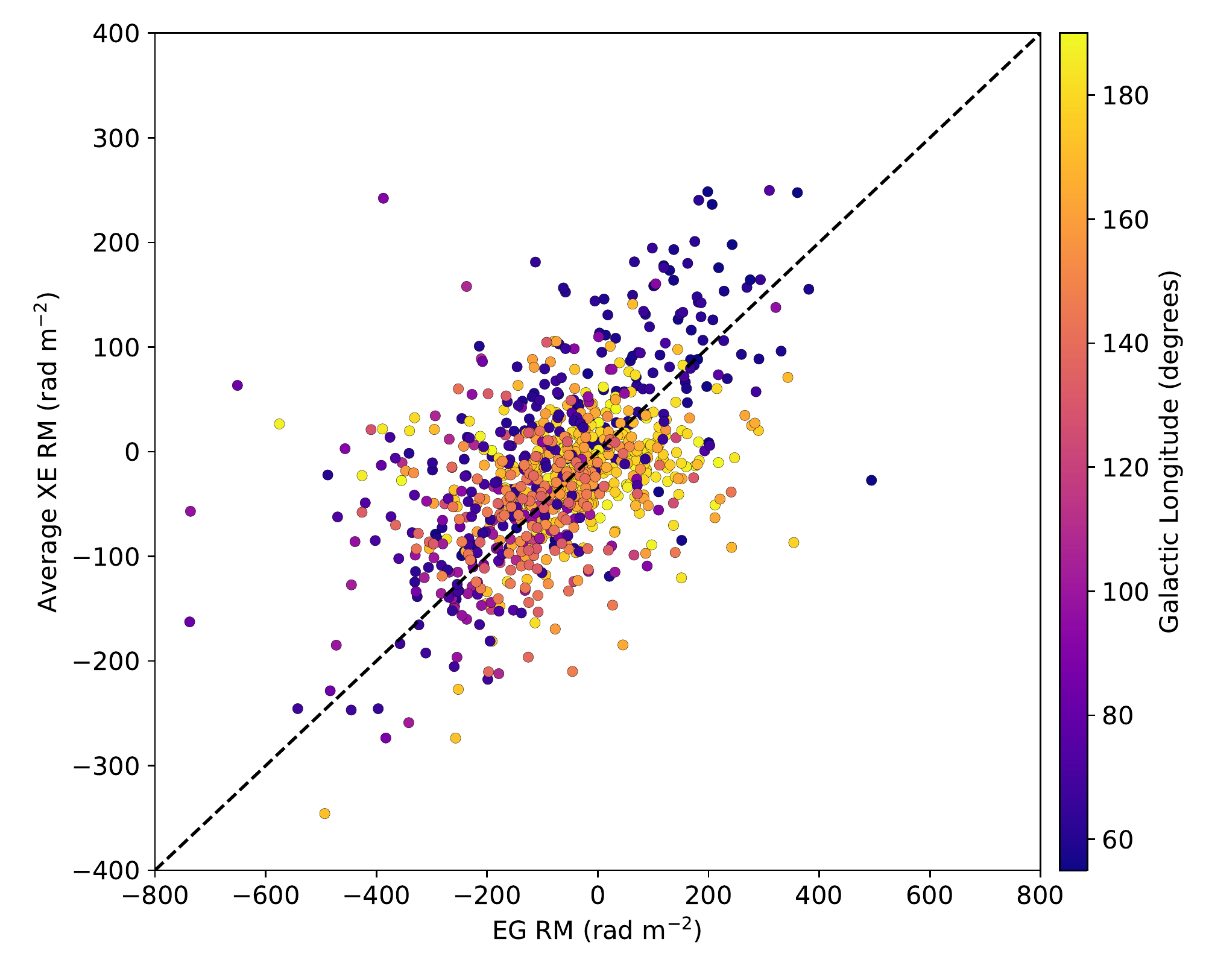}
   \caption{\added{The XE RM values averaged within circles with radius 0.25$^{\circ}$ centred at each EG source, plotted as a function of the corresponding EG RM. The colour scale indicates the locations of the sources in terms of Galactic longitude, and the dashed line indicates a constant ratio of 0.5 between the XE and EG RMs. The points are clustered close to this line, highlighting the correlation between the two types of sources.}}
\label{fig:scatterplot}
    \end{figure}

\added{To further examine the correlation between the XE and EG RMs, we also calculated average XE RM values within circles centred around each EG source. Varying the size of the circles, we calculated the resulting correlation coefficients, and found that the correlation is maximized for a radius of 0.25$^{\circ}$, corresponding to an area of approximately 0.20 square degrees. The result is shown in Figure~\ref{fig:correlation}. Using the size of the XE regions that maximizes the correletion, we plot the average XE RMs as a function of the corresponding EG RMs, as shown in Figure~\ref{fig:scatterplot}. We indicate the ratio of 0.5 between the two types of sources by the dashed line. Although there is a significant degree of scatter, the points are clustered around this line, with a root-mean-square difference of 80 rad m$^{-2}$. The highest degree of scatter occurs for sources at low longitudes, the reasons for which are discussed in Section \ref{s4.3}. }


\unskip

\added{It is important to note that unlike in Figure~\ref{fig:averaging} which utilizes independent bins, the calculated XE RM values in Figure~\ref{fig:scatterplot} overlap with some adjacent points due to the proximity of their locations, which are defined by the locations of the EG sources. This may, in part, account for the correlation between XE and EG sources being highest for XE regions significantly smaller than the independent 1.5$^{\circ}\times8^{\circ}$ bins, which result in an equally robust correlation. Furthermore, due to the variability of the EG source RMs as well (see Figure~\ref{fig:histograms}), averaging both types of sources (Figure~\ref{fig:averaging}) rather than treating the EG sources independently (Figure~\ref{fig:scatterplot}) allows for large-scale GMF structures to be discerned more easily. Regardless, Figure~\ref{fig:scatterplot} does highlight that the lines of sight corresponding to a significant number of EG sources pass through regions of XE with average RMs of approximately half the EG RM values.}

\section{Discussion}\label{s4}
Here we comment on the interpretation of the results described above. In particular, we discuss the relationship between the polarisation angles and the RM values, and present possible configurations of ISM parameters that may lead to the observed patterns. Additionally, we comment on regions in this dataset where the agreement between the XE and EG RMs breaks down.

\subsection{Polarisation Angles in the XE Data}
In addition to analyzing the RM values of the XE pixels, we investigated the corresponding observed polarisation angles at 1420 MHz \added{(not unwrapped to zero wavelength)}. Applying the \replaced{same filters to the map of polarisation angles as were applied to the RM map}{ S:N filter}, we found an approximately uniform distribution of the XE angles within most of the \added{longitude} bins. \added{The right panel of Figure~\ref{fig:histograms} shows an example of a bin within which the XE and EG RMs have clear peaks, but the polarisation angle distribution does not.} Less than 25\% of the bins demonstrated a weak peak in the polarisation angle histograms. These bins will require further investigation beyond the scope of this paper, but the lack of a dominant angle is likely due to the missing single-antenna data that would be required in order to detect large-scale structures in angle. Despite the absence of a coherent pattern in the distribution of angles, the \added{binned} RMs calculated from those angles\deleted{do have a meaningful average value in each bin, and those bin averages} vary smoothly and systematically with longitude, indicating that the XE are definitely reliable probes of the large-scale magnetic field in terms of RM values. The fact that small-scale fluctuations in the polarisation angle \added{at 1420 MHz} do not have a significant effect on bin-averaged RM values with our chosen bin size\deleted{, 1.5 degrees of longitude and 8 degrees of latitude,} is a surprising and useful~result.

\subsection{ISM Configurations}\label{s4.2}
\deleted{The results in} Figure~\ref{fig:averaging} show the strong resemblance between EG and XE RMs across a significant range of longitudes. This \replaced{similarity}{ immediately} suggests that the two types of sources \replaced{probe}{ are probing} the same volume of the ISM, and may put into question the previous assumption that XE RM data probe only the very local ISM (e.g.,~\cite{Uyaniker}). If XE and EG probe to a similar depth, this has important implications for interpreting polarisation data \replaced{in order to study}{ for the purpose of studying} the GMF. It may mean that \replaced{interpreting}{ the interpretation of} a Polarisation Horizon~\cite{Uyaniker} as a boundary beyond which contributions to the observed polarised intensity are insignificant, \replaced{does not apply}{ cannot be applied} in the same way to the XE RMs. We investigate this\deleted{idea} by considering\deleted{different} possible \added{ISM} configurations\deleted{of the ISM} that could give rise to the observed relationship between the XE and EG~RMs.

As described in the previous section, the ratio of EG source to XE source RMs is close to 2 for a large portion of the CGPS. The simplest configuration for the XE is the Burn slab model, for which the ratio of EG to XE RMs would indeed be 2\deleted{, but since the Galaxy is more complicated than a Burn slab, we have explored other options}. In the Burn slab model, synchrotron emission and Faraday rotation occur everywhere along the LOS through a region of uniform magnetic field, thermal electron density and synchrotron emissivity. \replaced{Since this is not accurate for the Galaxy}{ In this section}, we explore the extent to which \replaced{some non-uniform}{ more realistic large-scale} models of ISM parameters can resemble a slab-like scenario.

\deleted{The Faraday depth profile for the Burn slab is of the form:}


\deleted{\noindent where $x$ is the LOS distance between the observer and the source. The values of $\phi(x)$ range from 0 rad m$^{-2}$ to $\phi_{max}$, which is the Faraday depth of the far side of the slab. Assuming uniform complex synchrotron emission, $\tilde{P_{\circ}}$, the resulting Faraday depth spectrum~\cite{Brentjens} is:}


\deleted{\noindent Applying a Fourier transform to this function from $\phi$ space to $\lambda^2$ space, we can calculate the change in observed polarisation angles:}

\deleted{which is the well-known result for the expected RM from a Burn slab configuration. The Faraday depths from each point along the LOS average to yield a RM that is half of the value of the RM of a compact (Faraday thin) source seen through the same volume.}

\begin{figure}[H]
	\centering
	\includegraphics[width=0.5\textwidth]{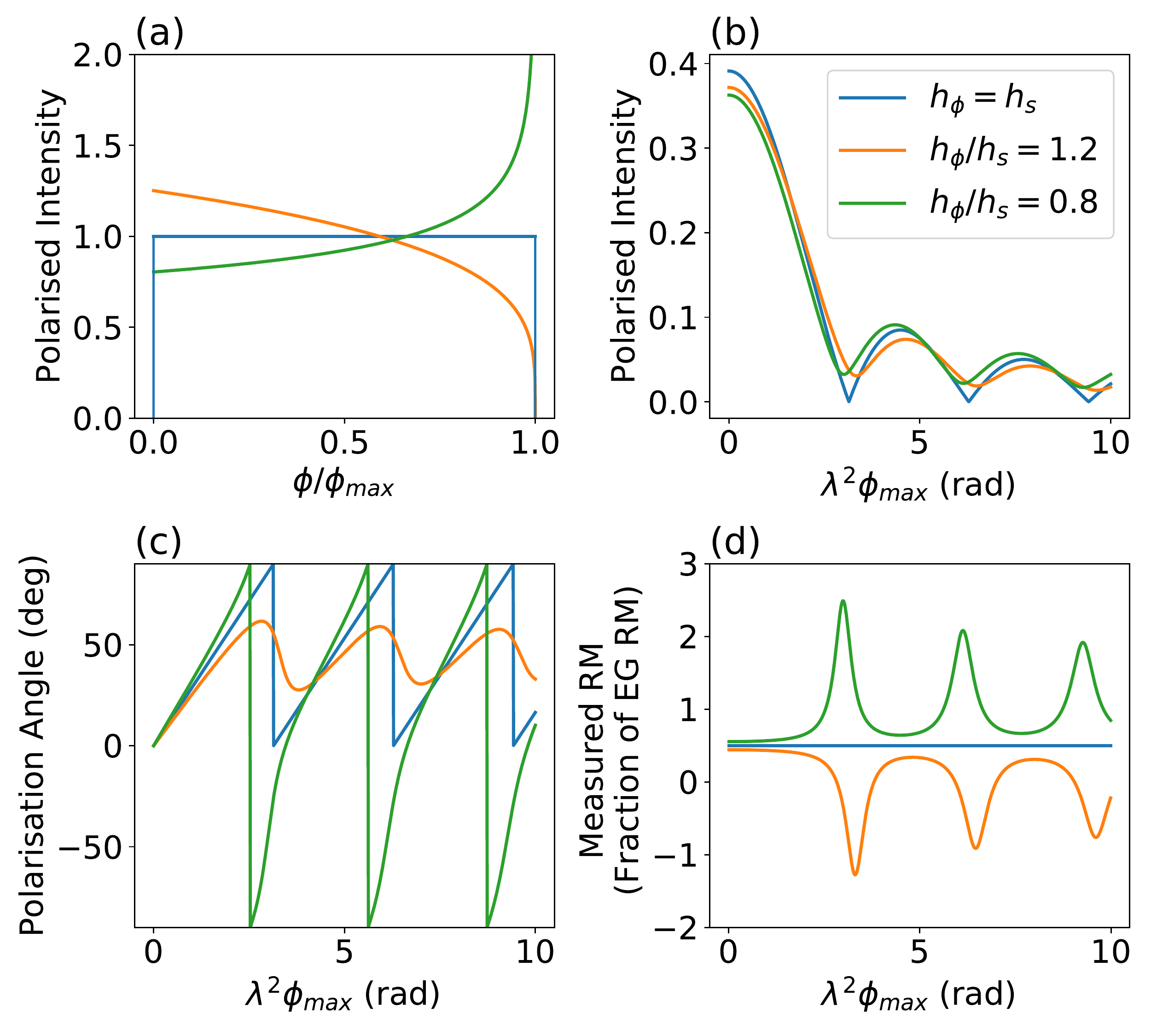}
	\caption{An example of a polarisation model, comparing the `slab-like' case (blue) where $h_{\phi}$ and $h_s$ are equal ($h_{\phi}/h_s=1$), to the case with $h_{\phi}/h_s=1.2$ (orange) \added{and the case with $h_{\phi}/h_s=0.8$ (green)}. (\textbf{a}): the Faraday depth spectrum. (\textbf{b}): polarised \replaced{intensity}{ flux} as a function of wavelength squared. (\textbf{c}): polarisation angle as a function of wavelength squared. (\textbf{d}): the slope of polarisation versus wavelength squared. Please note that in panels (\textbf{a}) and (\textbf{b}) the units of polarised \replaced{intensity}{ flux} are arbitrary and the curves have been normalized to have the same integrated \replaced{intensity}{ flux}.}
	\label{modelfig}
\end{figure}

For the interstellar medium, the thermal electron density, magnetic field strength and synchrotron emissivity can be reasonably modelled on large-scales as exponentials with characteristic scale-lengths:
$h_{n_e}$ \added{(electron density)}, $h_B$ \added{(magnetic field)} and $h_s$ \added{(emissivity).}\deleted{are the electron density, magnetic field and synchrotron emissivity scale-lengths respectively.} \added{In the simple model we describe here, the origin of the coordinate system for the exponential functions is located at the observer (i.e., the Earth) and not at the Galactic centre. Although this limits the model to lines of sight close to $\ell=180^{\circ}$, in this way it is possible to determine the Faraday depth spectrum by means of a straightforward analytical calculation.} The resulting Faraday depth spectrum\footnote{We use the tilde to denote a complex phasor encoding both the polarised intensity and polarisation angle.} (derived in the Appendix) is:
\begin{equation}
\tilde{P}(\phi)=\frac{\tilde{P}_{\circ}}{0.812n_{\circ}B_{\circ}}\Big(1-\frac{\phi}{\phi_{max}}\Big)^{\frac{h_{\phi}}{h_s}-1},
\label{FD_exp}
\end{equation}

\noindent where $\tilde{P}_{\circ}$, $n_{\circ}$ and $B_{\circ}$ are the polarised emission, electron density and magnetic field strength at the origin, $h_{\phi}=(1/h_{n_e}+1/h_B)^{-1}$ is the Faraday rotation scale-length, and $\phi_{max}=0.812n_{\circ}B_{\circ}h_{\phi}$ is the maximum possible Faraday depth in this configuration. We determine the Fourier transform numerically, and calculate observed polarisation angles as a function of wavelength squared, as shown in Figure~\ref{modelfig}. This derivation is similar to the `asymmetric slab' scenario presented by Sokoloff et al. 1998~\cite{Sokoloff}. Our approach \replaced{offers new insight}{ differs} in that we formulate the equations in terms of physical scale-lengths, calculate the Faraday depth spectra, and explicitly highlight three different regimes: (i) $h_{\phi}=h_s$; (ii) $h_{\phi}>h_s$; (iii) $h_{\phi}<h_s$.\\

\noindent Regime (i): $h_{\phi}=h_s$\\
\noindent \replaced{If}{ For the case where} Faraday rotation (\replaced{magnetic field and thermal electrons}{ characterised by thermal electron density and magnetic field scale-lengths}) and synchrotron emissivity (\replaced{magnetic field and cosmic ray electrons}{ characterised by the cosmic ray electron density and magnetic field scale-lengths}) extend the same distance out along the LOS, the polarised \replaced{emission}{ flux} has equal contribution from all Faraday depths out to $\phi_{max}$.\deleted{We can expect this result to hold for any distribution of emissivity and Faraday rotation, not necessarily an exponential, so long as the corresponding parameters are matched.}  In this case, Equation~(\ref{FD_exp}) simplifies to the Burn slab Faraday depth spectrum\deleted{of Equation~(\ref{FD_slab})}. \added{We can expect this result to hold for any distribution of emissivity and Faraday rotation, not necessarily an exponential, so long as the corresponding parameters are matched.}\deleted{It is noteworthy that this particular configuration of non-uniform ISM parameters is indistinguishable from a Burn-slab scenario in terms of the Faraday depth spectrum (Figure~\ref{modelfig}, (a)), the polarised \replaced{intensity}{ flux} as a function of $\lambda^2$ (Figure~\ref{modelfig}, (b)) and the polarisation angle as a function of $\lambda^2$ (Figure~\ref{modelfig}, (c)). Please note that in (Figure~\ref{modelfig}, (b)-(d)), $\lambda^2$ is parameterized as $\lambda^2 \phi_{max}$, to make it scale free. Consequently, the slope of the graph in (Figure~\ref{modelfig}, (c)) is a constant 0.5, the ratio between the detected XE RM and the maximum Faraday depth (Figure~\ref{modelfig}, (d)).} We compare the other two regimes to this slab-like regime. Henceforth, we use the term ``slab-like'' to indicate an ISM configuration that results in the same Faraday depth spectrum as the Burn-slab, but the parameters of which do not necessarily have uniform profiles.  \\

\noindent Regime (ii): $h_{\phi}>h_s$\\
\replaced{If}{ For the case where} Faraday rotation extends further out along the LOS than\deleted{the} synchrotron emissivity, the polarised \replaced{emission}{ flux} is concentrated at low Faraday depths \replaced{(nearby regions)}{, corresponding to more nearby regions (Figure~\ref{modelfig}, (a))}. \added{Unlike for the Burn slab,} the \replaced{polarised intensity versus}{ flux as a function of} $\lambda^2$\deleted{deviates slightly from a Burn-slab and} does not have null-points\deleted{(Figure~\ref{modelfig}, (b))}. The corresponding observed polarisation angles \replaced{versus}{ as a function of} $\lambda^2$ differ from the slab-like case in that the slope alternates between positive and negative values\deleted{(Figure~\ref{modelfig}, (c) \& (d))}. \replaced{In the limit as}{ The closer the ratio of} $h_{\phi}/h_s$ \replaced{approaches}{ is to} 1, \replaced{this approaches}{ the more this scenario resembles} a slab-like \replaced{case}{ model}: the Faraday depth spectrum flattens out, and the ranges of $\lambda^2$ over which the RM has a reversed sign decrease in extent.\deleted{Consequently, it will be statistically more likely that an observed value of XE RM will be close to half of the value of an EG RM for the same LOS, for a given range of detected wavelengths.}\\

\noindent Regime (iii): $h_{\phi}<h_s$\\
\replaced{When}{ For the case where} synchrotron emissivity extends \replaced{beyond the Faraday rotating region}{ further out along the LOS than Faraday rotation}, the polarised \replaced{emission}{ flux} is concentrated at higher Faraday depths\deleted{corresponding to} (more distant regions)\deleted{(Figure~\ref{modelfig}, (a))}. The \replaced{polarised intensity versus}{ flux as a function of} $\lambda^2$\deleted{(Figure~\ref{modelfig}, (b))} deviates from a Burn slab similarly to\deleted{the previous regime} \added{Regime (ii)}\deleted{(Figure~\ref{modelfig}, (b)). With $h_{phi} < h_s$}, \replaced{but there is an enhancement of the slope of polarisation angle versus $\lambda^2$ over some ranges of $\lambda^2$ rather than reversals in the sign.}{ the deviation from a slab-like model manifests itself not as reversals in the sign of polarisation angle versus $\lambda^2$, but as a strong enhancement of the slope over particular ranges of $\lambda^2$ (Figure~\ref{modelfig}, (c) \& (d)).} This leads to the surprising result that the observed XE RM\added{s}\deleted{values} can \replaced{exceed}{ actually be \textit{higher} than} the observed EG RM\added{s}\deleted{values} for certain wavelengths.\deleted{Once again, in the limit that the ratio $h_{\phi}/h_s$ approaches 1, this reduces to the slab-like model.}\\

Various models for the GMF have been tested that incorporate exponential profiles for the cosmic ray electron density and the magnetic field strength. Examples include the disk magnetic field model by Jaffe et al.~\cite{Jaffe}, in which the scale-length\added{s} of both the magnetic field and the cosmic ray electron density are 15 kpc, and the 3-dimensional models by Sun et al.~\cite{Sun}, in which magnetic field scale-lengths of 6 kpc, 8.5 kpc and 10 kpc are tested, while the cosmic ray electron density scale-length is 8 kpc . Both of these models use the NE2001 thermal electron density model~\cite{Cordes}. A 5 kpc scale-length is a reasonable estimate for an exponential fit to the NE2001 model. With this approximation, an estimate for the ratio of $h_{\phi}/h_s$ would be 0.8 for the Jaffe et al.~\cite{Jaffe} model, and 1.1 for the Sun et al.~\cite{Sun} model with a magnetic field scale-length of 8.5 kpc. These values are sufficiently close to 1 to allow for a ratio between XE and EG RMs of close to 0.5 over significant ranges of wavelengths.

\deleted{In the simple model we have described here, the origin of the coordinate system for the exponential functions is located at the observer (i.e., the Earth) and not at the Galactic centre. In this way, it was possible to determine the Faraday depth spectrum by means of a straightforward analytical calculation (See Appendix).}Assuming Galacto-centric exponential models for the parameters discussed, the spectrum we present is only strictly valid for the LOS directed toward $\ell=180^{\circ}$. For lines of sight with $\ell<180^{\circ}$, a greater portion of the LOS is dominated by the local ISM. For lines of sight with $\ell<90^{\circ}$, the emission and rotation profiles no longer decrease monotonically with LOS distance, which further complicates the model. In the scenario of a high Faraday depth dominated regime ($h_{\phi}<h_s$), there will be more emission at smaller Faraday depths in the lower longitude regions than at $\ell=180^{\circ}$. The effect of this increased emission at lower Faraday depths would be a shift toward a more slab-like regime for lower longitudes in the case of $h_{\phi}<h_s$ (predicted by the parameters in~\cite{Jaffe}). On the other hand, there would be a shift further from a slab-like regime for lower longitudes in the case of $h_{\phi}>h_s$ (predicted by the parameters in~\cite{Sun}).

\added{The model presented here is rudimentary and is intended as a toy model only. Furthermore, it is not possible to constrain the parameters using a dataset with limited frequency coverage. Nevertheless, the model is illustrative of the fact that it is possible to construct configurations of ISM parameters that result in a constant ratio of observed XE to EG RMs. Future work will include refining these models and incorporating more realistic aspects to explore exactly how the results would deviate from the basic structure described here. Another possibility for the ratio of 2 between EG and XE RMs is an averaging effect arising from the synchrotron emission being concentrated in a small number of patches with minimal depth, distributed randomly throughout the Galaxy.}

\subsection{Regions of Disparity between the XE and EG RMs}\label{s4.3}

While there is remarkable agreement between the average EG RMs and twice the average XE RMs over a significant portion of the CGPS span (as seen in Figure~\ref{fig:averaging}), there are a few noteworthy regions where we see disagreement between the two sources. Since these differences highlight interesting Galactic features, we discuss them here.

\subsubsection*{Latitudinal variation in HII complex: $\ell=172^{\circ}$}
At $\ell=172^{\circ}$ the XE RM $\times 2$ is seen to deviate significantly from the EG RM curve, and here both the XE and EG RMs suddenly become more negative than at longitudes $\pm 5^{\circ}$ from this location.  For both types of sources, RM values with such large magnitudes are unexpected as this longitude is near $\ell=180^{\circ}$, where the LOS becomes perpendicular to the field lines and as a result we expect RM values that are near zero.

\begin{figure}[H]
	\centering
    \includegraphics[width=1.0\textwidth]{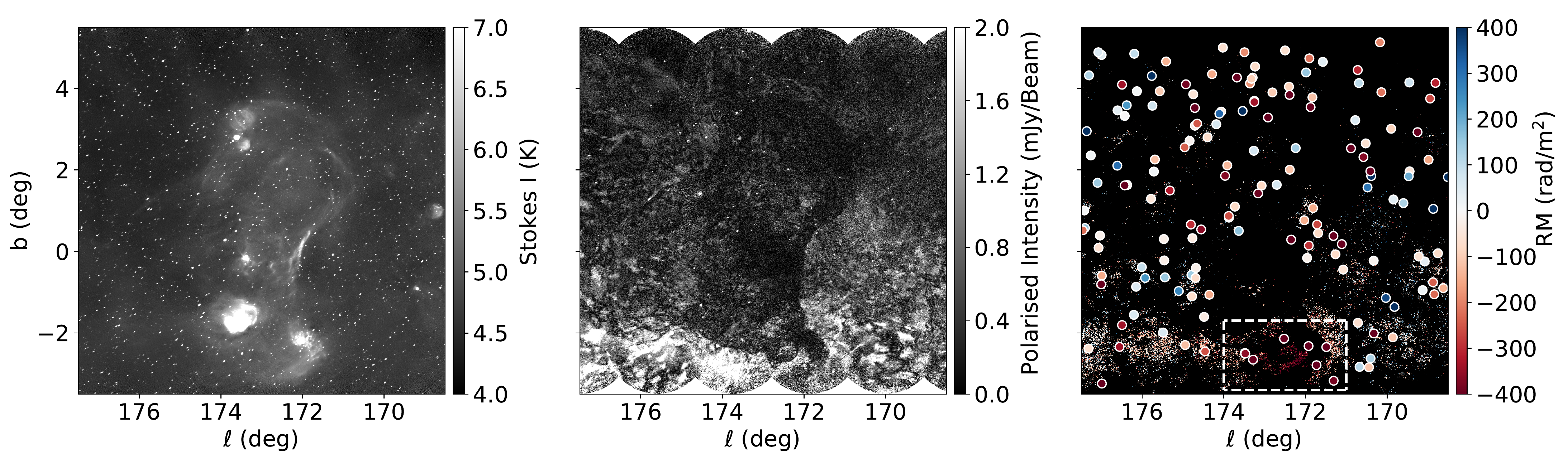}
   \caption{An HII region complex with an anomalous ratio of EG to XE RM values. \textbf{Left}: Stokes $I$ total intensity map. \textbf{Middle}: Polarised intensity map, revealing the depolarisation in this region. \textbf{Right}: RM values of EG sources (dots) and the XE (background pixels.) The dashed line box highlights a grouping of strongly negative EG RM sources.}
\label{fig:hourglass}
    \end{figure}

Closer inspection of the region near $\ell=172^{\circ}$, reveals numerous HII regions at a distance of about 2 kpc~\cite{Kang}, organized into a structure described by Gao et al.~\cite{Gao} as ``bow-tie''-like, as seen in the Stokes $I$ image in the left panel of Figure~\ref{fig:hourglass}. This structure corresponds to a depolarised bubble, as shown in the polarised intensity image in the middle panel of Figure~\ref{fig:hourglass}. As a result, there is very little XE data in this region, although there is a high density of EG sources. The exception to this is near the bottom of the structure, where we see unusually large RM magnitudes for both EG sources and XE. This is a Faraday screen estimated to be nearer than 300 pc~\cite{Gao}. Since the majority of the XE RM pixels in this longitude slice come from this small patch, the average XE RM magnitude in the bin at $\ell=172^{\circ}$ is considerably higher than the average EG RM magnitude, which results from sources at both high and low latitudes within this slice.

\unskip

\subsubsection*{Cygnus X region: $\ell=75^{\circ}$ to $85^{\circ}$}
Cygnus X is one of the most complex emission regions in the Galaxy. It contains hundreds of HII regions, stellar clusters and OB associations. There are three emission zones along the LOS at distances of 500--800 pc, at 1.7 kpc (a massive OB association), and other HII regions spread over 1.5 to 2 kpc~\cite{Gottschalk}. Together, these regions generate a high intensity of radio emission that is mostly thermal, making Cygnus X extremely bright in Stokes I, and relatively faint in polarised emission.

In the Cygnus X region there is a series of very high magnitude negative RM EG sources found at all latitudes. The strong negative dip in RM of EG sources is likely due to their emission passing though regions of high electron density in Cygnus X, enhancing their RM values. Polarised XE observed for these longitudes must have a very local origin, since Cygnus X has very strong depolarizing effects on more distant emission. Furthermore, nearly all of the XE pixels are discarded in the filtering process, due to the brightness of Cygnus X in Stokes $I$, leaving a small sample size of XE RMs. The result is that in Figure~\ref{fig:averaging} we see a significantly stronger negative average EG RM in the bins near $\ell=80^{\circ}$ than the average XE RM.

\added{The region near $\ell=108^{\circ}$ also has extremely low number of XE pixels remaining after filtering. Likely as a consequence of this, the XE and EG RMs deviate significantly at this longitude. Furthermore, the low number of XE pixels does not allow any meaningful interpretation of this deviation. }

\subsubsection*{Reversal region: $\ell=55^{\circ}$ to $\ell=70^{\circ}$}
An interesting feature of the XE and EG RM trends with longitude (Figure~\ref{fig:averaging}) is that the XE RMs appear to change sign at a slightly higher longitude (close to $\ell=67^{\circ}$) than the EG RMs (close to $\ell=60^{\circ}$). One possible explanation for this effect \replaced{may be the following.}{ is that,} If the depth to which the XE RMs probe is limited by a nearby polarisation horizon in this direction, then it is possible to have the XE RMs dominated by a\deleted{reversed} field region located relatively nearby, while the EG RMs for the same LOS are dominated by the field direction on the far side of the reversed region. At lower longitudes, where a greater portion of the EG source LOS passes through the reversed region, the EG source RMs will also reverse signs.

An alternative interpretation does not rely on the concept of a polarisation horizon, but rather is an extension on the slab-like models we presented in Section \ref{s4.2}. By including a reversal of the field in those models, we have found that as a function of the distance to the reversal, the expected ratio between the EG and XE RMs varies, and can be negative for some distances and positive for others. Since the distance to the boundary of the reversed region does vary with longitude, it is possible to construct configurations of the ISM parameters that lead to the observed result of the XE and EG RMs reversing sign at slightly different longitudes. Constraining the possible parameters for this is beyond the scope of this paper, but would be an interesting avenue for further investigation.

Considering the diagonal orientation of the boundary across which the reversal occurs, it may be argued that it is artificial to average over vertical bins in the longitude-latitude maps. Averaging the XE and EG RMs in bins parallel to the boundary was illustrated in Figure 3 of~\cite{Ordog}. However, calculating the ratios of those diagonally binned EG and XE RM values still does not yield a constant value in this region. This may be due to the possibility suggested in~\cite{Ordog} that the current sheet defining the boundary between opposing magnetic field directions is tilted not only in the plane of the sky-projection, but also along the line of sight, leading to complicated three-dimensional variations in the ratio of the detected EG to XE RMs near the reversal.

\section{Summary}\label{s5}

We have analyzed RMs of the polarised extended synchrotron emission (XE) along lines of sight through the Galactic disk, and compared these to RMs of extragalactic (EG) sources that are more commonly used as probes of the GMF. Our aim was to evaluate the reliability of the XE RMs calculated from the Synthesis Telescope polarisation data of the CGPS as a tool to probe the GMF. Potential concerns with this particular dataset may be the lack of single-antenna data in the 4-band observations, which leads to reduced sensitivity to large-scale structure in the \replaced{polarisation}{ polarised} images, and the limitations in detecting Faraday complexity that arise from using only four frequency channels. We have demonstrated that despite these potential limitations, there does appear to be useful information in this dataset.

We applied stringent filtering criteria to the XE RMs to exclude: (i) pixels that correspond to compact sources or lines of sight passing through extended structures that are not part of the diffuse emission, (ii) pixels with low signal to noise, and (iii) pixels with a poor linear fit to polarisation angle versus wavelength squared. Although this filtering process excluded a significant fraction of the available data, the remaining XE data still probe many more lines of sight than the EG sources.

To compare the CGPS XE and EG RMs, we plotted them as a function of Galactic longitude and found that over a large portion of the CGPS range they appear to follow the same pattern. This is promising, because the CGPS EG sources match the trends observed in other datasets, and have been determined to be a reliable tool for constraining models of the GMF. The fact that the XE sources also track the same trend lends credibility to this type of dataset being used for the same purpose.

Interestingly, the EG RM magnitudes are greater than the XE RM magnitudes by approximately a factor of 2 over a significant portion of the longitude range studied, which would be expected in the case of a Burn slab model of the ISM. A Burn slab model is unlikely to be realistic for most regions in the ISM, where the Faraday rotation and synchrotron emissivity vary along the line of sight. However, we have shown that as long as these parameters vary over similar scales, the resulting Faraday depth profile may be close to uniform, with the ratio between EG and XE RMs close to 2 over most wavelengths.

One of the regions where the ratio of EG to XE RMs deviates from 2 is near the reversal in the GMF. We have observed that the change in sign of the XE and EG RMs does not occur at the same longitude. This could be explained by a polarisation horizon in this region limiting the depth to which the XE probes the ISM, or by the asymmetric slab models we explored, combined with the distance to the reversal varying with longitude.

Based on our results, we conclude that the XE RMs are indeed a useful dataset, particularly for tracing the GMF in the outer Galaxy, $\ell>90^{\circ}$. The lines of sight at lower longitudes, probing a portion of the inner Galaxy, are more complicated due to the apparent large-scale reversal in the GMF in this region. For a comparison between the EG and XE sources in this region, the uncertainty in the interpretation of their individual RMs is compounded. At least for the outer Galaxy, the XE RMs can be incorporated into models of the GMF in combination with the more typically used compact source RMs. Improvements to this technique could be made by using combined single-dish and synthesis telescope data over a wider frequency range, in order to allow for added sensitivity to large-scale spatial structures and Faraday complexity along the lines of sight.

\vspace{6pt}



\authorcontributions{Conceptualization, A.O. and J.C.B.; methodology, A.O., R.A.B., C.V.E. and J.C.B.; formal analysis, A.O., R.A.B. and C.V.E.; investigation, A.O., R.A.B. and C.V.E.; writing—original draft preparation, A.O. and R.A.B.; writing—review and editing, C.V.E., J.C.B. and T.L.L.; supervision, J.C.B. and T.L.L.; funding acquisition, J.C.B. and T.L.L.}

\funding{The Canadian Galactic Plane Survey was a Canadian project with international partners. It was supported by the Natural Sciences and Engineering Research Council. The Dominion Radio Astrophysical Observatory is a national facility operated by the National Research Council Canada. This research has been supported by grants from the Natural Sciences and Engineering Research Council.}

\acknowledgments{We thank Bryan Gaensler and Jennifer West for helpful discussions about the data and ISM~models. We also thank the two reviewers whose comments and suggestions have helped to improve this manuscript.}

\conflictsofinterest{The authors declare no conflict of interest. The funding sponsors had no role in the design of the study; in the collection, analyses, or interpretation of data; in the writing of the manuscript, or in the decision to publish the results.}


\noindent
\begin{tabular}{@{}ll}

\end{tabular}

\appendixtitles{no} 
\appendixsections{one} 
\appendix
\section{}

Here we present the details of the analytical \replaced{model}{ calculation} outlined in Section \ref{s4.2} for determining the Faraday Depth spectrum corresponding to exponential models of ISM parameters. \added{For the Burn slab scenario, the Faraday depth profile is of the form:}
\begin{equation}
\phi(x)=0.812n_eB_{||}x,
\label{A1}
\end{equation}

\added{\noindent where $n_e$ is the electron density, $B_{||}$ is the LOS magnetic field component, and $x$ is the LOS distance between the observer and a specific location within the emitting slab. The values of $\phi(x)$ range from 0 rad m$^{-2}$ to $\phi_{max}$, which is the Faraday depth of the far side of the slab. Assuming uniform complex synchrotron emission, $\tilde{P_{\circ}}$, the resulting Faraday depth spectrum~\cite{Brentjens} is:}
\begin{equation}
\tilde{P}(\phi)=\begin{cases}
\mathLarge{\frac{\tilde{P}_{\circ}}{0.812n_e B_{||}}} &0\leq \phi \leq \phi_{max} \\
0 &\text{otherwise}.
\end{cases}
\label{A2}
\end{equation}

\added{\noindent Applying a Fourier transform to this function from $\phi$ space to $\lambda^2$ space, we can calculate the change in observed polarisation angles:}
\begin{equation}
\Delta \tau(\lambda^2)=\frac{1}{2}\lambda^2 \phi_{max},
\label{A3}
\end{equation}
\added{which is the well-known result for the expected RM from a Burn slab configuration~\cite{Burn, Sokoloff}. The Faraday depths from each point along the LOS average to yield a RM that is half of the value of the RM of a compact (Faraday thin) source seen through the same volume.}

\added{For the non-uniform models presented in Section \ref{s4.2},} we begin with exponential functions with different scale-lengths for the thermal electron density, magnetic field strength and synchrotron~emissivity:
\begin{eqnarray}
n_e(x)&=&n_{\circ}e^{-x/h_n}\nonumber \\
B_{||}(x)&=&B_{\circ}e^{-x/h_B}\nonumber \\
\tilde{P}(x)&=&\tilde{P}_{\circ}e^{-x/h_s}.
\label{A4}
\end{eqnarray}
From the electron density and magnetic field strength, we determine a Faraday Depth profile along the line of sight:

\begin{eqnarray}
\phi(x)&=&0.812\int_0^x n_{\circ}e^{-x/h_n} B_{\circ}e^{-x/h_B} dx\nonumber \\
&=&0.812n_{\circ}B_{\circ}\int_0^xe^{-\frac{x}{h_{\phi}}}dx\nonumber \\
&=&0.812n_{\circ}B_{\circ}h_{\phi}(1-e^{-\frac{x}{h_{\phi}}})\nonumber \\
&=&\phi_{max}(1-e^{-\frac{x}{h_{\phi}}}),
\label{A5}
\end{eqnarray}
where we have combined the scale-length parameters as $h_{\phi}=(1/h_{n_e}+1/h_B)^{-1}$. We define $\phi_{max}=0.812n_{\circ}B_{\circ}h_{\phi}$, which is the maximum possible Faraday rotation from this configuration. In general, the Faraday Depth spectrum can be calculated from the emission profile as:
\begin{equation}
\tilde{P}(\phi)=\int_0^{\infty}\delta (\phi-\phi(x))\tilde{P}(x)dx=\int_{-\infty}^{\infty}\delta (\phi-\phi(x))\tilde{P}(x)dx,
\label{A6}
\end{equation}
which corresponds to summing up the polarised \replaced{emission}{ flux} from all physical depths that result in each particular Faraday depth, $\phi$, thus building up a spectrum. Please note that the integral can be written with either 0 or $-\infty$ as the lower bound since we only detect \replaced{emission}{ flux} from one direction at a time.
\noindent The delta function can be re-written as:
\begin{equation}
\delta(\phi-\phi(x))=\sum\limits_{i=1}^n \frac{\delta(x-x_i)}{|\frac{\partial(\phi-\phi(x))}{\partial x}|_{x_i}}=\sum\limits_{i=1}^n \frac{\delta(x-x_i)}{|\frac{\partial\phi}{\partial x}|_{x_i}},
\label{A7}
\end{equation}
\noindent where the $x_i$ are the roots of its argument:
\begin{equation}
\phi-\phi(x_i)=0,
\label{A8}
\end{equation}
\noindent and $n$ is the number of roots.
\noindent Combining Equations~(\ref{A6}) and Equation~(\ref{A7}) results in:
\begin{equation}
\tilde{P}(\phi)=\int_{-\infty}^{\infty}\sum\limits_{i=1}^n\frac{\delta(x-x_i)}{|\frac{\partial\phi}{\partial x}|_{x_i}}\tilde{P}(x)dx=\sum\limits_{i=1}^n\frac{\tilde{P}(x_i)}{|\frac{\partial\phi}{\partial x}|_{x_i}}.
\label{A9}
\end{equation}
\noindent The integral can be dropped because the integrand only picks up the locations where $x=x_i$. For the Faraday Depth profile described by Equation~(\ref{A5}), Equation~(\ref{A8}) has only one root:
\begin{eqnarray}
\phi(x_i)&=&\phi=\phi_{max}(1-e^{-\frac{x_i}{h_{\phi}}})\nonumber\\
x_i&=&-h_{\phi}\text{ln}\Big(1-\frac{\phi}{\phi_{max}}\Big).
\label{A10}
\end{eqnarray}
The derivative evaluated at the root is:
\begin{eqnarray}
\Big|\frac{\partial \phi}{\partial x}\Big|_{x_i}&=&\frac{\phi_{max}}{h_{\phi}}e^{-\frac{x_i}{h_{\phi}}}\nonumber \\
&=&\frac{\phi_{max}}{h_{\phi}}\Big(1-\frac{\phi}{\phi_{max}}\Big).
\label{A11}
\end{eqnarray}
Combining Equation~(\ref{A9}) with Equation~(\ref{A10}) and Equation~(\ref{A11}), results in:
\begin{eqnarray}
\tilde{P}(\phi)&=&\frac{\tilde{P}_{\circ}e^{-x_i/h_s}}{\frac{\phi_{max}}{h_{\phi}}\Big(1-\frac{\phi}{\phi_{max}}\Big)}\nonumber \\
&=&\frac{\tilde{P}_{\circ}e^{\frac{h_{\phi}}{h_s}\text{ln}\Big(1-\frac{\phi}{\phi_{max}}\Big)}}{\frac{\phi_{max}}{h_{\phi}}\Big(1-\frac{\phi}{\phi_{max}}\Big)}\nonumber \\
&=&\frac{h_{\phi}\tilde{P}_{\circ}}{\phi_{max}}\Big(1-\frac{\phi}{\phi_{max}}\Big)^{\frac{h_{\phi}}{h_s}-1}\nonumber \\
&=&\frac{\tilde{P}_{\circ}}{0.812n_{\circ}B_{\circ}}\Big(1-\frac{\phi}{\phi_{max}}\Big)^{\frac{h_{\phi}}{h_s}-1}.
\label{A12}
\end{eqnarray}
Applying a Fourier transform to this Faraday depth spectrum allows us to determine the detected polarisation as a function of wavelength squared that results from this configuration:
\begin{equation}
\tilde{P}(\lambda^2)=\int_{-\infty}^{\infty}\tilde{P}(\phi)e^{2i\lambda^2\phi} d\phi=\int_0^{\phi_{max}}\tilde{P}(\phi)e^{2i\lambda^2\phi} d\phi.
\label{A13}
\end{equation}
We evaluated this Fourier transform numerically for different values of the ratio $h_{\phi}/h_s$ to yield the results shown in Figure~\ref{modelfig}.



\reftitle{References}





\end{document}